\documentclass[letter,longauth,traditabstract,printer]{aa}  
\usepackage{graphicx}
\usepackage{txfonts}
\usepackage{natbib}
\bibpunct{(}{)}{;}{a}{}{,} 

\newcommand{\micron}{$\mu$m}

\begin{document}
   \title{Resolving debris discs in the far-infrared: early highlights from the DEBRIS survey\thanks{{\it Herschel} is an
ESA space observatory with science instruments provided by
European-led Principal Investigator consortia and with important
participation from NASA.}}

   \author{B.C. Matthews
          \inst{1,2}
          \and
	  B. Sibthorpe\inst{3},
	  G. Kennedy\inst{4},
	  N. Phillips\inst{5},
	  L. Churcher\inst{4},
	  G. Duch\^{e}ne\inst{6,7},
	  J.S. Greaves\inst{8},
	  J.-F. Lestrade\inst{9},
	  A. Moro-Martin\inst{10,11},
	  M.C.~Wyatt\inst{4},
	  P. Bastien\inst{12}
	  A. Biggs\inst{3},
	  J. Bouvier\inst{6},
	  H.M.\ Butner\inst{13},
	  W.R.F.\ Dent\inst{14},
	  J. Di Francesco\inst{1,2},
	  J. Eisl\"{o}ffel\inst{15},
	  J. Graham\inst{7},
	  P. Harvey\inst{16},
	  P. Hauschildt\inst{17},
	  W.S.\ Holland\inst{3},
	  J. Horner\inst{19},
	  E. Ibar\inst{3},
	  R.J.\ Ivison\inst{3,5},
	  D. Johnstone\inst{1,2},
	  P. Kalas\inst{7},
	  J. Kavelaars\inst{1,2},
	  D. Rodriguez\inst{20},
	  S. Udry\inst{21},
	  P. van der Werf\inst{22},
	  D. Wilner\inst{23},
	  B. Zuckerman\inst{20}
          }

   \institute{Herzberg Institute of Astrophysics, National Research Council Canada, 5071 West Saanich
              Road., Victoria, BC, Canada, V9E 2E7 \\
              \email{brenda.matthews@nrc-cnrc.gc.ca} 
         \and
	 University of Victoria, Finnerty Road, Victoria, BC, V8W 3P6
              Canada \\
	 \and
	 	 UK Astronomy Technology Center, Royal Observatory, Blackford
              Hill, Edinburgh EH9 3HJ, UK \\
	 \and
	 Institute of Astronomy, University of Cambridge, Madingley
  Road, Cambridge, CB3 0HA, UK \\
	      \and
	      Institute for Astronomy, University of
Edinburgh, Royal Observatory, Blackford Hill, Edinburgh EH9 3HJ,
              U.K. \\
              \and
	      Laboratoire d'Astrophysique, Observatoire de Grenoble,
              Universit\'{e} J. Fourier, CNRS, France \\
	      \and
	      Department of Astronomy, University of California, 601
              Campbell Hall, Berkeley, CA, U.S.A., 94720 \\
	      \and
School of Physics and Astronomy, University of St Andrews,
  North Haugh, St Andrews, Fife KY16 9SS, UK \\
  \and
  Observatoire de Paris - CNRS, 77 Av. Denfert Rochereau, 75014 Paris,
              France \\
	      \and
Centro de Astrobiolog\'{i}a (CSIC-INTA), 28850 Torrej\'{o}n de Ardoz,
              Madrid, Spain \\
	      \and
	      Department of Astrophysical Sciences, Ivy Lane, Peyton
Hall, Princeton University, Princeton NJ 08544 USA \\
              \and
	      D\'{e}partement de physique et Observatoire du
              Mont-M\'{e}gantic, Universit\'{e} de Montr\'{e}al,
              C. P. 6128, Succ. Centre-ville, Montr\'{e}al, QC H3C
              3J7, Canada \\
	      \and
	      Department of Physics and Astronomy, James Madison University, Harrisonburg, VA 22807, U.S.A \\
              \and
	      ALMA JAO,  Avda. Apoquindo 3846, Piso 19, Edificio
              Alsacia, Las Condes, Santiago, Chile \\
	      \and
	      Th\"{u}ringer Landessternwarte, Sternwarte 5, D-07778
              Tautenburg, Germany \\
	      \and
	      Astronomy Department, University of Texas at Austin, 1
              University Station C1400, Austin, TX 78712-0259,
              U.S.A. \\
	      \and
	      Hamburger Sternwarte, Gojenbergsweg 112, 21029 Hamburg,
              Germany \\
	      \and
	      Department of Physics and Astronomy, The Open
	      University, Milton Keynes MK7 6AA, U.K. \\
	      \and
	      Department of Physics, Science Laboratories, Durham
              University, South Road, Durham, DH1 3LE, U.K. \\
	      \and
	      Dept. of Physics \& Astronomy, University of California, Los Angeles, 475
              Portola Plaza, Los Angeles, CA 90095-1547, USA \\
	      \and
	      Geneva Observatory, Astronomy Department of the Geneva
              University, Switzerland \\
	      \and
	      Leiden Observatory, Leiden University, Postbus 9513,
              2300 RA, Leiden, The Netherlands \\
	      \and
	      Harvard-Smithsonian Center for Astrophysics, 60 Garden Street, Cambridge, MA 02138
             }

   \date{Received March 31, 2010;  accepted May 9, 2010}

\offprints{B.~Matthews}

\titlerunning{Resolving debris discs in the far-infrared with {\it Herschel}}

\authorrunning{B.~Matthews et al.}

 \abstract{We present results from the earliest observations of
   DEBRIS, a {\it Herschel} Key Programme to conduct a volume- and
   flux-limited survey for debris discs in A-type through M-type
   stars.  PACS images (from chop/nod or scan-mode observations) at
   100 and 160 \micron\ are presented toward two A-type stars and one
   F-type star: $\beta$ Leo, $\beta$ UMa and $\eta$ Corvi. All three
   stars are known disc hosts.  {\it Herschel} spatially resolves the
   dust emission around all three stars (marginally, in the case of
   $\beta$ UMa), providing new information about discs as close as 11
   pc with sizes comparable to that of the Solar System.  We have
   combined these data with existing flux density measurements of the
   discs to refine the SEDs and derive estimates of the fractional
   luminosities, temperatures and radii of the discs.  }


   \maketitle
%

\section{Introduction}
\label{intro}

Debris discs are flattened distributions of planetesimals and dust
located at radii of 1-1000~AU around main-sequence stars (see
\citep[see][for a recent review]{wyatt08}. The dust cannot be
primordial since its lifetime in orbit is significantly less than the
age of the host stars.  Instead, dust is replenished from a population
of colliding km-sized planetesimals \citep{wyatt02,the07}.  Over time,
the dust distribution is shaped by any planetary-sized bodies in the
system \citep[e.g.,][]{dom03,wyatt07}. Therefore, resolved images of
discs constrain models of the structure and evolution of planetary
systems.

Far-infrared and submillimetre observations are the best way to search
for dust around nearby stars due to the favorable contrast of the disc
relative to the star.  At these wavelengths, the disc emission is optically
thin and is sensitive to the large (up to $\sim 1$ mm), cold grains which
dominate the disc's dust mass.

The {\em Herschel} Space Observatory offers three major advantages for
the detection and characterization of debris discs: far-infrared
sensitivity, angular resolution and wavelength coverage. With its 3.5
m mirror, its sensitivity at far-infrared wavelengths is superior to
any previous instrument. With its resolution of 6\farcs7 at 100
\micron, {\it Herschel} has the potential to resolve many debris
discs, particularly toward nearby stars. Finally, with detectors at
100, 160, 250, 350 and 500 \micron, {\it Herschel} has the means to
sample the spectral energy distribution (SED) of disc emission across
the peak, meaning models can be better constrained even for discs
which are not resolved.

DEBRIS (Disc Emission via a Bias-free Reconnaissance in the
Infrared/Submillimetre) is an Open Time Key Programme which uses PACS
(Photodetector Array Camera \& Spectrometer) and (for appropriate
targets) SPIRE (Spectral and Photometric Imaging Receiver) to detect,
resolve and characterize debris discs around a volume-limited sample
of 446 A through M type stars.  The goals of DEBRIS include
establishing the incidence and evolution of debris discs as a function
of stellar type, age, multiplicity, etc.; the characterization of
discs in terms of size, temperature, dust mass and morphology (where
the disc asymmetries could indicate the presence of planetary
companions); and the understanding of our own Solar System in the
context of the larger debris disc population.  Full details of the
DEBRIS survey and goals will be presented in a forthcoming paper
(B.~Matthews et al. 2010, in preparation).  Here, we present PACS
observations toward three of the first targets of the DEBRIS survey.
We briefly summarize the observations and targets in Sect.~\ref{obs},
present the results in Sect.~\ref{res} and discuss three sources in
detail in Sect.~\ref{disc}. We summarize the paper in Sect.~\ref{sum}.


\begin{figure*}
\centering
\includegraphics[bb=20 335 640 645,width=20cm,clip]{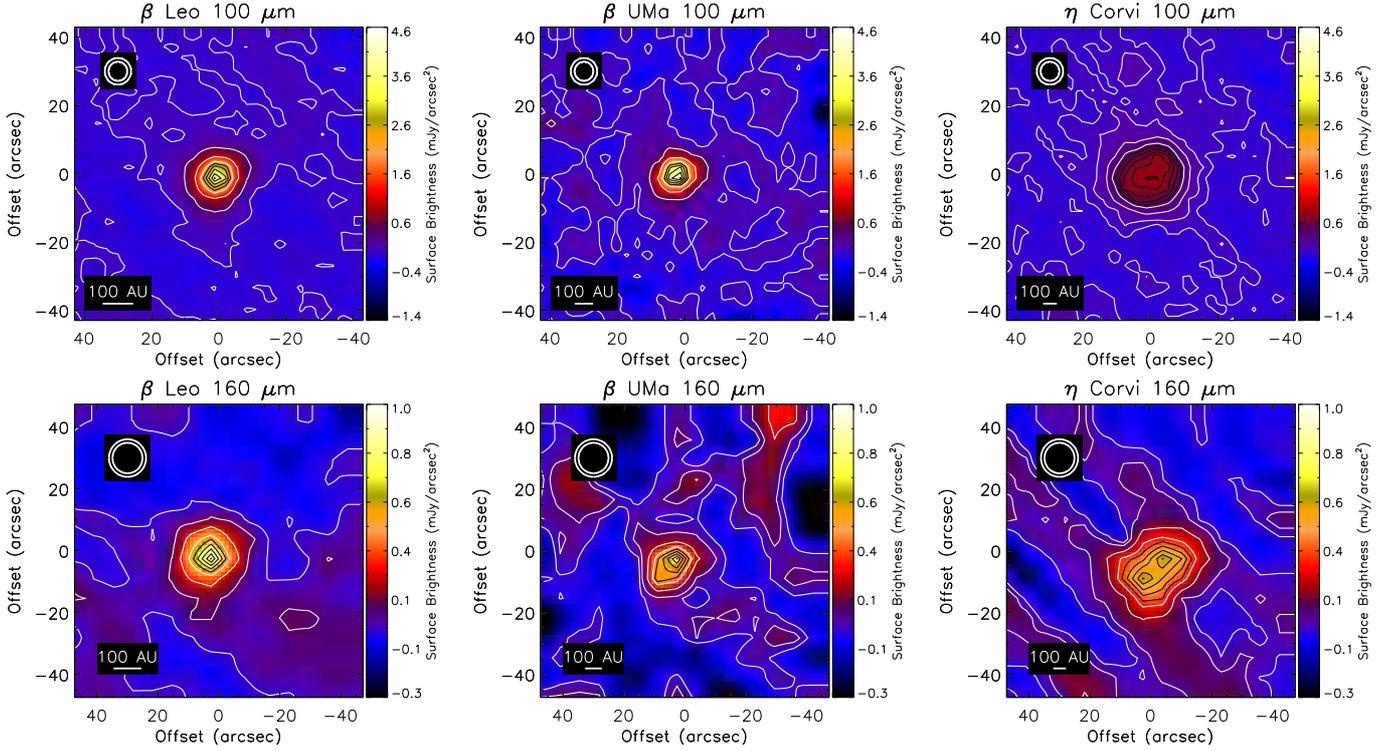}
\caption{Images of the 100 and 160 \micron\ emission from three DEBRIS
  targets: $\beta$ Leo, $\beta$ UMa and $\eta$ Corvi. Contours are
  shown at 0, 10, 30, 50, 60, 70, 80, 90 and 99\% of the peak in each
  map. The 1-$\sigma$ rms noise levels are given in Table \ref{table}.
  Circles in the upper left corner of each panel mark the nominal beam
  sizes for a scan speed of 20\arcsec/s, i.e., 6\farcs7 and 11\arcsec\
  at 100 and 160 \micron, respectively. The negative images created by
  the chop/nod observing mode are visible in the $\beta$ UMa
  images. Striping in the $\eta$ Corvi image at 160 \micron\ is due to
  high 1/f noise and filtering artefacts.  }
\label{images}
\end{figure*}

\section{Observations and data reduction}
\label{obs}

DEBRIS is a flux-limited survey and as such it observes each target to
a uniform depth (1.2 mJy beam$^{-1}$ at 100 \micron), resulting in
different mass limits for targets at different distances and of
different stellar spectral types. Here, we present 100 and 160
\micron\ photometry observations toward three nearby stars (see Table
\ref{table}) performed with the ESA {\it Herschel} Space Observatory
\citep{pil10} utilizing the PACS \citep{pog10} instrument.  The
results presented here were taken during early testing phases or
during the Science Demonstration Phase on {\it Herschel} (2009 Sept.\
- Dec.)

The images were obtained using two different observing strategies:
point source chop/nod, and small scan-map modes (see the PACS
Observers'
Manual\footnote{http://herschel.esac.esa.int/Docs/PACS/html/pacs\_om.html}).
For point-source mode observations, seven contiguous repeat chop/nod
observations were performed. Scan map observations had eight repeats
in a single scan direction at a rate of 20\arcsec/s.  Four
3\arcmin\ scan legs were performed per map with a 2\arcsec\ separation
between legs.  The total observing times were 1072 and 1220 seconds,
respectively, for each chop/nod and scanning observation.

\begin{figure}[h!]
\includegraphics[scale=0.8]{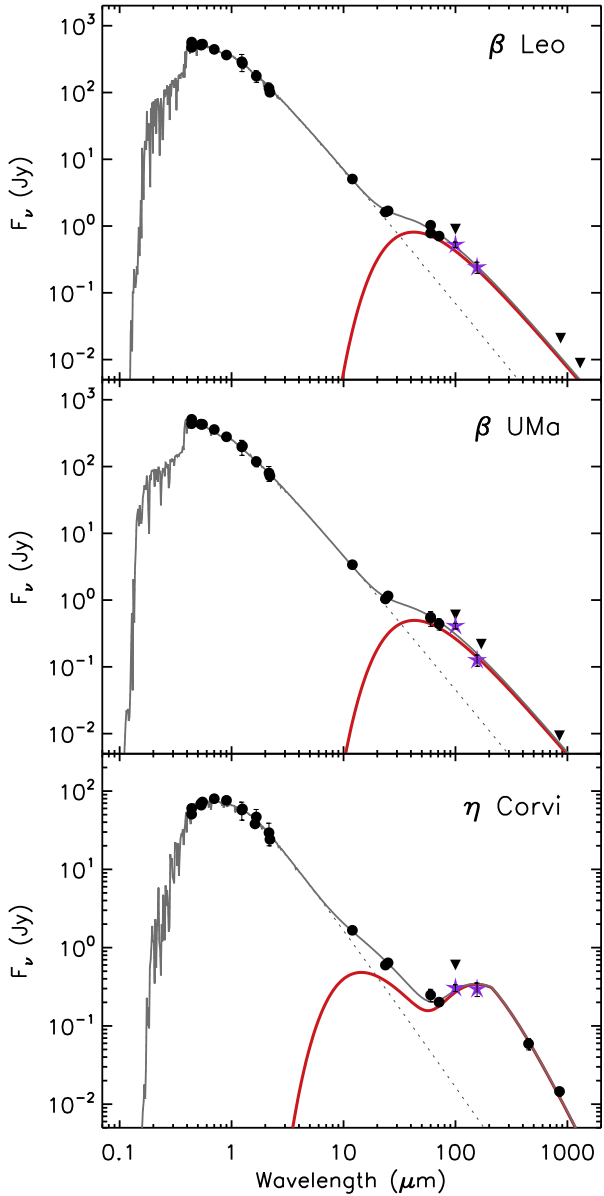}
\caption{Spectral energy distributions of three DEBRIS targets (grey
  lines): $\beta$ Leo, $\beta$ UMa and $\eta$ Corvi (top to
  bottom). Stars mark the 100 and 160 \micron\ flux densities from
  {\it Herschel}.  Optical data were obtained from Simbad and
  Vizier. {\it Spitzer} 24 and 70 \micron\ data are from \cite{su06}
  and \cite{beichm06}. IRAS 60 and 100 \micron\ flux densities are
  reprocessed through SCANPI. ISO fluxes are from \cite{hab01}.
  Submillimetre photometry for $\beta$ Leo and $\eta$ Corvi are from
  \cite{holmes03} and \cite{wyatt05}, respectively.  The submillimetre
  upper limit for $\beta$ UMa is from the SCUBA (Submillimetre Common
  User Bolometer Array) archive. We use synthetic photometry to fit
  NextGen model atmospheres (dotted lines) assuming $\log(g)=4.5$,
  ${\rm [M/H]} = 0$, and no reddening. Blackbody disc models (thick
  red lines) are fit to star-subtracted excesses following
  \cite{rhee07}.  We use a modified blackbody with $210/\lambda$
  beyond 210 \micron\ for the cold component of $\eta$ Corvi,
  reflecting the inefficient emission from grains at (sub-)mm
  wavelengths \citep{wyatt08}.  The star and disc parameters are
  presented in Table \ref{table}.}
\label{SEDs}
\end{figure}

Table \ref{table} shows the survey (``UNS'') identifier for each
target as well as the source and observing details.
\cite{phil09} contains details of the development of the Unbiased
Nearby Stars sample from which the DEBRIS targets are drawn.
  
These data were reduced using the {\it Herschel} interactive processing
environment \citep[HIPE][]{ott10}.  Maps were obtained via the default PACS na\"ive
map-making methods \emph{photProject} and
\emph{photProjectPointSource} in HIPE for the scanning and point source
observing modes respectively.  Scanned data were pre-filtered to
remove low frequency (1/\emph{f}) noise using a boxcar filter with
width equal to 1\farcm5.  All bright sources in the map were masked
prior to filtering to avoid filter ringing type artefacts.  The
chop/nod configuration meant that no equivalent filtering was required
for the data obtained in point source mode.

All three targets presented in this paper are shared targets with the
DUNES {\it Herschel} Key Programme (PI: C.\ Eiroa) which has science
goals complementary to those of DEBRIS.  Details on the distribution
of targets will be discussed in a survey description paper
(B.~Matthews 2010, in preparation).

\begin{table}
\caption{Stellar and disc parameters}
\label{table}
\centering
\begin{tabular}{lccc}
\hline\hline
Parameter & $\beta$ Leo & $\beta$ UMa & $\eta$ Corvi \\
\hline
UNS ID & A005 & A024 & F063 \\
HD number &  HD 102647 & HD 95418 & HD 109085 \\
Spec. Type & A3 Va & A1 V & F2 V \\
PACS OM & scan & point-source & scan \\
D [pc] & 11.0 & 24.3 & 18.2 \\
rms$_{100}$ [mJy/beam] & 1.4 & 3.1 & 1.2 \\
rms$_{160}$ [mJy/beam] & 3.9 & 7.6 & 5.0  \\
Aperture [arcsec] & 20 & 15 & 20 \\
$F_{100}$ [mJy] & $500\pm 50$ & $390 \pm 39$ & $300\pm 30$ \\
$F_{160}$ [mJy] & $230\pm 46$ & $120\pm 25$ & $290\pm 58$ \\
\hline
$\rm{FWHM}^{100}_{min}$ [\arcsec]  & $9.2 \pm 0.1$
  & $7.2 \pm 0.1$ & $16.4 \pm 0.4$  \\
$\rm{FWHM}^{100}_{maj}$ [\arcsec] & $10.4 \pm 0.1$ & $8.4
  \pm 0.2$ & $18.3 \pm 0.4$ \\
$PA^{100}$ [$^\circ$] & $125 \pm 3$ & $114 \pm 5 $ & $102 \pm 7$  \\
$L^*$ [$L_\odot$] & 13.5 & 60 & 5.0 \\
$T^*_{eff}$ [K] & 8380 & 9340 & 6950  \\
$f_D = L_{IR}/L^*$ & $2.2 \times 10^{-5}$ & $1.4 \times 10^{-5}$ & $3.6\times 10^{-4}$  \\
$T_{disc}$ [K] & 112 & 109 & 31, 354 \\
$R_{dust}$ [AU] & 23 & 51 & 174, 1.4 \\
$R_{obs}$\tablefootmark{a} [AU] & $\sim 39$ & $\sim 47$ &
  $\sim 145$ \\
\hline
\end{tabular}
\tablefoottext{a}{Estimates of radius are deconvolved from the beam,
  assuming $R_{obs}/D =$ sqrt(FWHM$^{100}_{min}$ $\times$
  FWHM$^{100}_{maj}$ $- BEAM^2$).}
\end{table}


\section{Results}
\label{res}

Figure \ref{images} shows the 100 and 160 \micron\ images for the three
targets.  The rms levels achieved in each observation are summarized
in Table \ref{table}. The higher noise levels associated with the
point-source mode are evident. The scan map noise levels were
significantly lower for comparable observing times and, for $\eta$
Corvi, meet the DEBRIS rms specifications. For $\beta$ Leo, the rms is
higher by $\sim$ 15\%.

The integrated flux densities are estimated for each image (star +
disc). This is done with simple aperture photometry using apertures
(see Table \ref{table}) centred on the peak emission.  \cite{pog10}
detail the flux calibration of PACS data and estimate the calibration
uncertainties in the measured flux densities to be 10\% and 20\% for
100 and 160 \micron, respectively. The dominant flux calibration
uncertainties have been combined in quadrature with statistical
uncertainties from the rms levels in the maps.  These combined
uncertainties are applied to the fluxes in Table \ref{table}.

The flux densities reported in Table \ref{table} are plotted on
spectral energy distributions in Fig.\ \ref{SEDs}.  The disc
components of $\beta$ Leo and $\beta$ UMa are well fit by a simple
blackbody in the absence of submillimetre detections, but $\eta$
Corvi requires a two component fit to its disc emission: a warmer
blackbody and a modified blackbody for the cold component to fit the
submillimetre flux densities.  The temperature, radius ($R_{dust}$)
and fractional luminosity of these fits are reported in Table
\ref{table}.

Fitting of 2D Gaussians to each source at 100 \micron\ yields FWHM
values (see Table \ref{table}) larger than the nominal PACS PSF
of 6\farcs7. Analysis of an observation of Vesta yields a PSF of
6\farcs6 $\times$ 6\farcs9. Vesta is a cool blackbody for which the
response within the 100 \micron\ filter should be very similar to our
dust discs.  It has a temperature measured in the
submillimetre at 130-160 K \citep{chamber07}, slightly warmer than our
two A-star dist discs at $\sim$110 K.  Taking into account the
spectral response theoretically, the range of FWHM varies by less than
5\%, for slopes from $-2$ to $+1$ in $\lambda \ F_{\lambda}$.  Larger
PSFs have been measured yielding maxiumum dimensions as high as
7\farcs3.  Since only the long axis of the $\beta$ UMa disc exceeds
this size, we claim this disc is marginally resolved. The $\beta$ Leo
and $\eta$ Corvi discs are well resolved at 100 \micron.  Estimates of
deconvolved disc radius have been made from the FWHM. We call this
radius estimate $R_{obs}$ (Table \ref{table}).

\section{Discussion}
\label{disc}

\subsection{$\beta$ Leo}
\label{betaleo}

Figure \ref{images} shows the first resolved images of the disc around
$\beta$ Leo.  The blackbody temperature and dust luminosity results
given in Table \ref{table} are consistent with the values found in
previous works \citep{su06,holmes03}.  The fractional dust luminosity
of $2.3 \times 10^{-5}$ is 15\% higher than the estimate from
\cite{su06}.  The increase is due to a slight increase in dust to
match the PACS flux densities.

The radius estimates found for $\beta$ Leo are comparable to that of
the Kuiper Belt ($\sim 50$ AU). This makes the $\beta$ Leo disc one of
the smallest disc radii yet resolved at any wavelength (see, for
instance, the 'Circumstellar Disks
Database'\footnote{cirumstellardisks.org}) although smaller
characteristic orbital radii have been derived based on single
temperature blackbody fits to the dust components
\citep[e.g.,][]{rhee07}.

The difference between $R_{obs}$ and $R_{dust}$ provides an
opportunity to learn about the grains within this disc.  Because
differently sized grains can have the same temperature at different
distances from a star, the SED models in Fig.\ \ref{SEDs} are
degenerate. This degeneracy is broken by the resolved imaging. For
example, the $\sim$40 AU radius for the $\beta$ Leo disc is larger
than the 23 AU suggested by the blackbody fit. Therefore, the
grains do not emit as blackbodies, but maintain a $\sim$112 K
temperature at a greater distance from the star as expected for small
grains that emit inefficiently at far-IR wavelengths.  The inferred
characteristic particle radius $a$ is $ < \lambda/2 \pi =
16$ \micron. Future modeling work that combines {\it Spitzer} IRS spectra and
submillimetre images with the {\it Herschel} data will constrain these grain
properties and the spatial dust distribution (L.~Churcher et al. 2010,
in preparation).

\subsection{$\beta$ UMa}
\label{betauma}

Figure \ref{images} shows that the disc emission around $\beta$ UMa is
very compact at 100 and 160 \micron.  The disc is marginally resolved
at 100 \micron\ and not resolved at 160 \micron. The apparent
asymmetry in the 160 \micron\ disc image is likely artificial; it is
an effect of interpolation applied to the image at native (Nyquist
sampled) resolution.  The flux densities measured for $\beta$ UMa
confirm the earlier 100 and 160 \micron\ detections.

The disc component of the $\beta$ UMa SED is well fit by blackbody
grains with a temperature comparable to that of $\beta$ Leo, requiring
a bigger disc around the more luminous star.  Therefore, assuming
black body grains the radial estimate is 51AU, equivalent to the
deconvolved disc radius from the 2D Gaussian fit to the 100 \micron\
image.  The resolved size thus suggests an absence of small
grains such as that inferred for beta Leo in Sect.~\ref{betaleo}.
More detailed modeling will be forthcoming in a future paper.

\subsection{$\eta$ Corvi}

The new {\it Herschel} images in Fig.\ \ref{images} show that $\eta$
Corvi is resolved at both 100 and 160 \micron, as expected based on
the $\sim 300$ AU submillimeter size derived by \cite{wyatt05}.  The
variation in morphology from centrally peaked emission at 100 \micron\
to a double-peaked limb brightened ring at 160 \micron\ (as observed
at 450 \micron) is consistent with an outer cool ring filled in by
warmer dust which dominates the emission at 100 \micron. This could be
evidence of the third temperature component proposed by \cite{chen06}
and observed in $\epsilon$ Eri by \cite{back09}, although this was
tentatively ruled out in mid-IR imaging by \cite{smith08}, and more
generally suggests the radial distribution of material is broader than
the two ring system originally envisaged by \cite{wyatt05}.

The $R_{obs}$ estimate from Table \ref{table} is equivalent to the
submillimetre size. The two intensity maxima in the 160 \micron\ image
are roughly a beamwidth (11\arcsec) apart, identical to the 450
\micron\ SCUBA imaging of \cite{wyatt05} who inferred that the
emission arises from a ring at moderate inclination.
Fitting a 2D Gaussian to the 100 \micron\ image of Fig.\ \ref{images}
gives a position angle of 102\degr $\pm 7$\degr and an inclination of
$\sim$ 50\degr\ from the line of sight. The position angle of the two
peaks at 160 \micron\ is $\sim 135$\degr. These inclinations are
consistent with the 450 \micron\ measurement of $130$\degr\ $\pm 10$\degr.

A two component model of the SED of Fig.\ \ref{SEDs} shows similar
results to \cite{wyatt05} who found disc components of 40 K and 370
K. The warm component \citep{smith09,smith08,chen06} shown in
Fig.\ \ref{SEDs} has a blackbody temperature of 346 K, corresponding
to a radial distance of 1.4 AU from the star.  The cold component has
a temperature of 33 K, corresponding to a radial separation of 160 AU
from the star, consistent with $R_{obs}$.  As for $\beta$ UMa,
the resolved size suggests an absence of the small grains implied for 
$\beta$ Leo in Sect.~\ref{betaleo}.  

Most importantly, the images in Fig.\ \ref{images} provide an estimate
of the disk size at wavelengths intermediate between the submillimetre
(which shows emission at $\sim 150$ AU) and mid-IR (which shows
emission at $<$ 3.5 AU). This will be crucial for modeling the origin
of the far-infrared morphology, which most resembles the submillimetre
emission.  Simultaneously modeling these several images will constrain
in more detail the dust properties of the disc system.

\section{Summary}
\label{sum}

These early images of known debris disc hosts highlight the resolving
power of {\it Herschel}. For $\eta$ Corvi, the addition of resolved
images in the FIR provides important constraints on the outer disc
grain properties, and our data support the presence of a warmer inner
component to the cool outer ring.  We have resolved the discs around
$\beta$ Leo and (marginally) $\beta$ UMa for the first time and find
that both have sizes on the order of the Kuiper Belt.  Both are among
the smallest discs yet resolved.

The DEBRIS project will push the detection limits for debris discs
around nearby stars towards Kuiper-Belt levels.  {\it Spitzer}
volume-limited surveys achieved an rms of about 5 mJy at 70
\micron\ \citep[e.g.,][]{trilling08}, compared to our 1.2 mJy rms at 100
\micron, where the contrast to the photosphere is generally also
increased.  This improved sensitivity, coupled with the improved
resolution of {\it Herschel} and when applied to the large sample of discs
that DEBRIS will ultimately observe, will satisfy the paucity of
direct measurements of disc sizes that currently impedes modelling of
debris discs.

\begin{acknowledgements}

We thank our referee, K.~Stapelfeldt, for a constructive and insightful
report. Support for this work, part of the NASA {\it Herschel} Science
Center Key Program Data Analysis Program, was provided by NASA through
a contract (No.\ 1353184, PI: H. M. Butner) issued by the Jet
Propulsion Laboratory, California Institute of Technology under
contract with NASA.  This project is supported by a Space Science
Enhancement Program grant from the Canadian Space Agency.

\end{acknowledgements}

\end{document}